\documentclass[onecolumn,preprintnumbers,amsmath,amssymb]{revtex4}
\usepackage{graphicx}
\begin{document}

\title{Noise-induced intermittence}

\author{David Lambert$^1$, Luigi Palatella$^2$, Paolo Grigolini$^1$ }
\address{$^1$Center for Nonlinear Science, University of North Texas,
   P.O. Box 311427, Denton, Texas 76203-1427\\
   $^2$Liceo Scientifico Statale ``C. De Giorgi'', viale De Pietro 14, I-73100 Lecce (Italy)}
  
\begin{abstract}
We study a form of noise-induced intermittence  originated by an out of equilibrium process yielding events in time with a survival probability that in the case of an infinitely aged condition coincides with the Mittag-Leffler function. In contrast with the Pomeau-Manneville intermittence, the aging process does not have any effect on the inverse power law of the large time scale but on the short-time stretched exponential regime.

\end{abstract}

\maketitle

\section{Introduction}

Intermittence is a phenomenon of central interest in the field of complexity as made evident from a sample of recent papers \cite{intermittence1,intermittence2,intermittence3,intermittence4,intermittence5,intermittence6}. The theoretical arguments adopted to discuss intermittence reflect a wide set of opinions and theoretical tools, ranging from the Continuous Time Random Walk (CTRW) perspective \cite{ctrw1,ctrw2,ctrw3} to Record Dynamics \cite{intermittence1,intermittence2}, from a beneficial use of the CTRW concepts \cite{intermittence6} to a strong criticism of CTRW \cite{intermittence1}, from the use of a multiplicative random-advection term yielding intermittency \cite{intermittence4} to the adoption of stochastic thermodynamics \cite{intermittence5}. The range of applications is very wide, too, ranging from evolutionary path \cite{intermittence1} to glassy dynamics \cite{intermittence3}, from Blinking Quantum Dots (BQD) \cite{bqd1,bqd2,bqd3} to the physics of liquid crystals \cite{liquidcrystals}. 

The  perspectives behind the theoretical results of this article are, (i) the \emph{weak chaos} foundation of intermittence and (ii) the \emph{noise-induced intermittence}. The weak chaos foundation of intermittence is based on the popular Pomeau-Manneville (PM) map \cite{pm}. As interesting examples of the former kind of intermittence we quote the recent papers of Refs. \cite{korabel1,korabel2,akimoto}.  The important connection between intermittence and invariant measure was discussed by Tasaki and Gaspard \cite{tasaki} and the authors of Refs. \cite{korabel1,korabel2} recover their main conclusion
that in the non-stationary regime these maps realize a non-normalizable invariant density. Although this important subject is still to some extent controversial (see Ref. \cite{brasilian}), we believe that there is a complete agreement about the connection between non-stationary and non-integrability of the invariant measure. We notice that the numerical calculations are done using a finite number of copies of the system (Gibbs ensemble) and consequently, having in mind the approach to the invariant measure proposed by the authors of Ref. \cite{rosa}, the condition of non-normalizable density corresponds to distribution density remaining in a perennial out of equilibrium condition.  In conclusion, in the literature there seems to be a substantial agreement about the fact that power law intermittence corresponds to a physical condition where the conventional invariant density cannot be realized.

To define the processes of noise-induced intermittence we refer to recent 
attempts at deriving from a dynamical model of interacting units the kind of intermittence revealed by the experiment on the physics of BQD yielded only a partial success \cite{physicstodayblinking,beyondkramers,partial1,partial2,adam}. 
We note that the illuminating discussion of Ref. \cite{physicstodayblinking} shows that the process of regression to the origin
in a merely diffusional process shares the renewal nature of the blinking quantum dots.  Margolin and Barkai, by properly revisiting the Kramers theory \cite{beyondkramers} pointed out that before the traditional exponential regime of the Kramers theory the process of regression to the origin is characterized by $
\mu = 1.5$. The authors of Ref. \cite{partial1} proposed a model of interacting units that at criticality generate a renewal non-Poisson intermittence, with a power index $\mu = 1.5$ and the authors of Ref. \cite{partial2} found a wider range of power indexes, which are based only on numerical calculations with no theoretical support. The more recent work of Svenkeson, Bologna and Grigolini \cite{adam} establishes the theoretical foundation of $\mu = 1.5$ by proving that the decision making model of Ref. \cite{partial1} generates the Langevin equation
\begin{equation} \label{adambologna}
\dot{x} = - \gamma x^3 + f(t).
\end{equation}
The intermittence observed by means of this model has to do with the regression to the origin of the variable $x$, which generates an inverse power law with $\mu = 1.5$, truncated by the action of the restoring potential, and so realizing the condition discussed by Margolin and Barkai \cite{beyondkramers}. Setting $\gamma = 0$ would have the effect of making the inverse power law infinitely extended, but the power index $\mu = 1.5$ is generated by free-diffusion, thereby making the theoretical prediction indistinguishable from that generated by a merely diffusion process \cite{bqd2}, in conflict with the intuitive argument that the BQD physics may have an anomalous origin. 

A form of noise-induced intermittence released from the ordinary diffusion constraint $\mu = 1.5$ is discussed in the recent work of Ref. \cite{luigi}. The authors of this paper proved that the adoption of a logarithmic potential, much weaker than the quartic potential of Eq. (\ref{adambologna}), generates regressions to the origin with no truncation. It has to be pointed out that in the processes of noise-induced intermittence noise is the cause of the inverse power law structure of the re-crossings to the origin, in a sharp contrast with the condition of intermittence generated by weak chaos, where noise is known instead to generate the exponential truncation of inverse power law intermittence \cite{bettin,elena}.

The main aim of this article is to establish a connection between  the noise-induced intermittence and the generalized exponential relaxation proposed many years ago by Mittag-Leffler, and now becoming increasingly popular, as proved by a sample of some interesting papers on this subject 
 \cite{ml1,ml2,ml3,ml4,ml5,ml6}.  Using this theoretical perspective we  
 find a restoring potential much weaker than the quartic potential of Eq. (\ref{adambologna}).  A good approximation assigns to this potential a logarithmic form, 
 thereby recovering the results of Ref. \cite{luigi}, with  noise-induced intermittence generating a power index $\mu$ that is not limited to the condition $\mu = 1.5$ but it extends from $\mu = 1.5$ to $\mu = 2$. It is remarkable that the index $\mu = 2$ does not require any approximation being the exact theoretical property of the ML theoretical approach in the limiting case of extremely slow ML decay.

   As earlier mentioned, renewal aging is an interesting property of intermittence generated by weak chaos. We show that noise-induced intermittence is associated to a different form of aging. We prove, in fact, that noise-induced intermittence is realized by events with a ML survival probability and we show that aging affects its short- rather than its long-time region.   

The outline of this article is as follows. In  Section \ref{ml} we show how to use the ML survival probability to derive a Langevin equation that is the generator of the noise-induced intermittence of this article. In Section \ref{pm} we illustrate the origin of aging in the case of intermittence generated by weak chaos. This is done for the main purpose of making as clear as possible the form of aging emerging from the model of this article. 
This new form of aging,  not affecting the long-time behavior of the survival probability, is discussed in Section \ref{newaging}. 
Note that the analytical results of this article are based on the logarithmic idealization, but Section \ref{david} addresses the difficult issue of going beyond the logarithmic idealization and establishes the key role of the condition $\mu = 2$ with analytical arguments. Finally Section \ref{final} is devoted to concluding remarks.

\section{The ML function as a  source of stochastic dynamics} \label{ml}
Let us consider the ML function
\begin{equation} \label{supers}
E_{\alpha}(-(\lambda t)^{\alpha}) = \int_{0}^{\infty} \mathrm{d}q \Pi(q) exp(-qt).
 \end{equation}
With a simple algebra based on the definition itself of ML function \cite{demonstration}, it is straightforward to obtain the normalized distribution density
\begin{equation} \label{normalizable}
\Pi(q) = \frac{q^{\alpha-1}\sin(\pi\alpha) \lambda^{\alpha}}{\alpha\left(q^{2\alpha} + 2(q\lambda)^{\alpha} \cos(\pi \alpha) + \lambda^{2\alpha}\right)}.
\end{equation}
We note that, in spite of diverging at $q = 0$ and decaying very slowly for $q \rightarrow \infty$, $\Pi(q)$ is integrable in both regions, thereby making 
it possible to interpret the ML survival probability of Eq. (\ref{supers}) according to the following picture. The parameter $q$, as in the earlier work of \cite{bianco}, is assumed to be a rate of event production, for example the rate of photon emission by a chromophore \cite{fluctuatingenzyme}. 
We consider a Gibbs ensemble corresponding to the distribution density $\Pi(q)$, and we let this ensemble produce events. In other words, the ML survival probability is the sum of infinitely many exponential relaxations, a condition reminiscent of superstatistics \cite{beck}. In super-statistics a non-canonical energy distribution is assumed to be the sum over infinitely many canonical distributions corresponding to different temperatures. Here the ML survival probability is the sum of infinitely many relaxation processes with different values of the rate of event production. 

It is important to point out that Eq. (\ref{supers}) implies that the observation of event production starts at a time when the Gibbs ensemble is at equilibrium.
In this article we prepare the Gibbs system in the condition corresponding to
a virtually vanishing value of $q$ and we have to find a prescription for the time evolution of $q$ that at very large times, when equilibrium is reached, may lead to an observation process corresponding to Eq. (\ref{supers}). 
Thus, we make the assumption that 
$\Pi(q)$ is the equilibrium distribution generated by the following non-linear Langevin equation:

\begin{equation} \label{langevin}
\dot q = - \frac{d}{dq}V(q) + f(t),
\end{equation}
where $f(t)$ is an uncorrelated noise with $\langle f \rangle = 0$ and $\langle f^2 \rangle > 0$.
We establish the analytical form that $V(q)$ must have to generate an equilibrium condition that coincides with the normalized distribution density $\Pi(q)$ of Eq. (\ref{normalizable}). This is done 
by setting the condition 
\begin{equation}
p_{eq}(q) = \frac{exp\left(-\frac{V(q)}{D}\right)}{Z} \propto \Pi(q). 
\end{equation}
We develop the theory of this article adopting dimensionless variables, thereby setting $\lambda = 1$,  replacing Eq. (\ref{supers}) with 
\begin{equation} \label{supers2}
E_{\alpha}\left(-t^{\alpha}\right) = \int_{0}^{\infty} \mathrm{d}q \Pi(q) e^{-qt}
\end{equation}
and $\Pi(q)$ of Eq. (\ref{normalizable}) with 
\begin{equation} \label{normalized2}
\Pi(q) = \frac{q^{\alpha-1}\sin(\pi \alpha)}{\alpha\left(q^{2\alpha} + 2 q^{\alpha} \cos(\pi \alpha) + 1\right)}.
\end{equation}

This leads us to the potential $V$ defined by 
\begin{equation} \label{potential}
\frac{d}{dq} V \equiv D \left[\frac{(\alpha -1)}{q} - \frac{2\alpha{}q^{\alpha-1}\left(q^{\alpha} + \cos(\pi \alpha) \right)}{q^{2\alpha} + 2 q^{\alpha} \cos(\pi \alpha) +1} \right]
\end{equation}
and to the following Langevin equation
\begin{equation} \label{fundamental1}
\dot q = D \left [\frac{(\alpha -1)}{q} - \frac{2\alpha{}q^{\alpha-1}\left(q^{\alpha} + \cos(\pi \alpha) \right)}{q^{2\alpha} + 2 q^{\alpha} \cos(\pi \alpha) +1}\right] +
\sqrt{2D} f(t) ,\end{equation}
where  $\langle f \rangle = 0$ and $\langle f^2 \rangle = 1$. This
corresponds to the Fokker-Planck equation
\begin{equation} \label{fundamental}
\frac{\partial}{\partial t} p(q,t) = D \frac{\partial}{\partial q} \left [\frac{(1- \alpha)}{q} + \frac{2\alpha{}q^{\alpha-1}\left(q^{\alpha} + \cos(\pi \alpha) \right)}{q^{2\alpha} + 2 q^{\alpha} \cos(\pi \alpha) +1}\right]p(q,t) + D \frac{\partial^2}{\partial q^2} p(q,t) .\end{equation}

Note that in the short-$q$ region the potential responsible for the motion of the variable $q$ becomes indistinguishable from 
\begin{equation} \label{short}
\frac{d}{dq} V(q) = D\frac{1 - \alpha}{q}.
\end{equation}
This approximation establishes a connection between the model of this article  and the logarithmic potentials that are currently the object of investigation by some researchers \cite{luigi,iddo,dechant,campisi}.

To establish a connection with the search for the intermittence of blinking quantum dots,  free from the constraint $\mu = 1.5$, we focus our attention on the recursion to the origin. It is important to notice that it is easy to extend the model to the case where $q$ may also be negative. We assume $V(-q) = V(q)$ and we interpret the time evolution with $q < 0$ as corresponding to the BQD dark state, the light state being represented by $q > 0$. Of course, this procedure coincides with assigning to the laminar regions between to consecutive arrivals to $q=0$ alternate signs.

We assume that the trajectory is initially located in a region around the origin of size $a$ and we plan to estimate the time distance between two consecutive regressions of the trajectory to this region. We expect that the waiting time distribution density $\psi^{(BQD)}(\tau)$ in one of the two BQD states has the inverse power law structure
\begin{equation}
\psi^{(BQD)}(\tau) \propto \frac{1}{\tau^{\mu}}.
\end{equation}
The population $p(x,t), x < a$ of this region depends on the regression to the origin according to the formula
\begin{equation} \label{regression}
p(x, t) \propto \sum_{n=1}^{\infty} \psi_n^{(BQD)}(t),
\end{equation}
where $\psi_n(t)$ is the probability density for the particle to be back to the region $x < a$ at time $t$ after $n-1$ earlier regressions. 
Assuming that the population of the region $x< a$ has the time dependence
\begin{equation} \label{important1}
p(x,t) \propto \frac{1}{t^{\delta}},
\end{equation}
we get
\begin{equation} \label{renewal}
\mu = 2 - \delta.
\end{equation}
To find $\delta$ we adopt the logarithmic approximation
\begin{equation} \label{ideal}
\dot q = - \frac{D(1-\alpha)}{q} + \sqrt{2D} f(t)
\end{equation}
and the corresponding Fokker-Planck equation
\begin{equation} \label{ideal2}
\frac{\partial}{\partial t} p(q,t) =  D \left[\frac{\partial}{\partial q} \frac{(1-\alpha)}{q} +  \frac{\partial^2}{\partial^2 q} \right]p(q,t). 
\end{equation}

We also set the reflecting boundary condition
\begin{equation}
\frac{d}{dq} p(q,t) = 0
\end{equation}
at $q = 0$.
To realize this condition we assume that in the short region $0 < q < a$ the potential vanishes and we set for the potential $V(q)$ as
\begin{equation}\label{forkramers}
V(q) = \left \{ 
\begin{array}{lll}
0 & \mathrm{for} & q < a\\
 D(1-\alpha)  \log \left(\dfrac{q}{a}\right)
 & \mathrm{for} & q \geq a\\
\end{array}
\right .
\end{equation}
This assumption makes our treatment exactly equivalent to the one adopted in the recent work of Ref. \cite{dechant}, which leads us to 
\begin{equation} \label{important2}
\delta = \frac{\alpha}{2}.
\end{equation}
Using Eq. (\ref{renewal}) we obtain 
\begin{equation} \label{palatella}
\mu = \frac{4-\alpha}{2}.
\end{equation}

 Notice that this result agrees with that obtained in the earlier work of Ref.\cite{luigi}. 
This is of some interest because in spite of the close similarity, the approach adopted in this paper is not identical to that of the earlier work of Ref. \cite{luigi}.  In \cite{luigi} the authors
study the pdf of the times a system following Eq.(\ref{ideal}) starting at $q>0$ 
needs to reach the origin where the expression (\ref{ideal}) is violated.
The authors find that the power law tail exponent $\mu$, following the notation
of Eq.(\ref{ideal}), reads
\begin{equation}
\mu = \dfrac{3}{2} + \dfrac{D(1-\alpha)}{\sqrt{(2D)^2}} = \dfrac{3}{2}
+ \dfrac{1 - \alpha}{2} = \dfrac{4 - \alpha}{2},
\end{equation}
yielding  the same of the same result  as  this paper.  Eq. (\ref{palatella}). Nevertheless it is important to stress
that this time the problem we are facing is  different, because in this paper we are studying the
waiting time between two consecutive returns to the origin of a system following Eq.(\ref{ideal})
for $q>a$ with a cutoff in the drift term for $q < a$ (see Eq. (\ref{forkramers})).

This is a remarkable result because it yields an intermittent process going beyond the diffusion restriction $\mu = 1.5$. 
 Eq. (\ref{forkramers}) shows that when  $\alpha = 1$ the restoring potential vanishes, making the time evolution of $q$ depend only on  the stochastic force $f(t)$, and thus explaining why in this case $\mu = 1.5$.  Furthermore, and this is an impressive difference with the condition of Eq. (\ref{adambologna}), the action of the restoring logarithmic potential does not generate any truncation of the inverse power law. It is important to stress that the quartic potential of Eq. (\ref{adambologna}) is produced by a cooperation strength of the cooperating units of the decision making model of Refs. \cite{partial1}  large enough as to produce criticality.  It would be very interesting to prove that the restoring potential $V(q)$ generating $\mu \neq 1.5$ is the effect of a still unknown phase transition that generates the logarithmic potential of Eq. (\ref{short}) rather than producing the quartic potential of Eq. (\ref{adambologna}).

The logarithmic approximation generates interesting results, but, as we shall see in Section \ref{david}, the prediction of Eq. (\ref{palatella}) at long times is violated. In the short-time regime the variable $q$ remains reasonably small so as to be compatible with the logarithmic approximation of Eq. (\ref{short}). In the long-time regime the variable $q$ explores a region of the potential $V$ where
the corrections to the approximation of Eq. (\ref{short}) become significant and the prediction of Eq. (\ref{palatella}) is violated. However, these corrections tend to vanish for $\alpha \rightarrow 0$ and the prediction of Eq. (\ref{palatella}) that 
$\mu = 2$ for $\alpha = 0$ turns out to be correct for both short and long times.

\section{Aging of the PM process} \label{pm}
We devote this Section to reviewing the renewal aging  of intermittence processes generated by weak chaos. The idealized form of PM map reads
\begin{equation}
\dot y = r y^z,
\end{equation}
with $z > 1$.
The trajectory $y(t)$ moves from the initial condition $y$ with
$0 < y < 1$ to $y(t) = 1$ in a time $\tau$ given by 
\begin{equation} \label{selecting}
\tau = \frac{1}{r} \left\{\frac{1}{y^{\frac{1}{\alpha} } - 1}\right\},
\end{equation}
where
\begin{equation}
\alpha = \frac{1}{z - 1}.
\end{equation}
The corresponding waiting time distribution of age $t_a$, $\psi(\tau, t_a)$, is derived from
\begin{equation} \label{general}
\psi(\tau, t_a) = p(y, t_a) |dy/d{\tau}|,
\end{equation} 
with $p(y,t_a)$ denoting the distribution density of the initial condition $y$ at time 
$t_a$ after the preparation of the initial condition $p(y, 0) = 1$. 
The time evolution of $p(y,t_a)$ is driven by \cite{rosa}
\begin{equation} \label{cheerful}
\frac{\partial}{\partial t_a} p(y,t_a) = - r \frac{\partial}{\partial y} y^z p(y,t_a) + \alpha p(1,t_a) .
\end{equation}
In the case where $z < 2$ it is straightforward to prove 
that the equilibrium distribution of Eq. (\ref{cheerful}) is
\begin{equation} \label{equilibrium}
p_{eq}(y) = p(y, t_a = \infty) = \frac{2-z}{y^{z-1}}.
\end{equation}

The brand new and the infinitely aged waiting time distributions are obtained
using Eq. (\ref{general}) with $p(y, t_a) = 1$ and  $p(y, t_a = \infty)$ given by Eq. (\ref{equilibrium}), respectively. With straigthforward calculations we get:
\begin{equation} \label{brandnew}
\psi(\tau, t_a = 0) = \alpha \frac{T^{\alpha}}{(\tau + T)^{\alpha +1}}
\end{equation}
and
\begin{equation} \label{old}
\psi(\tau, t_a = \infty) = (\alpha -1) \frac{T^{\alpha-1}}{(\tau + T)^{\alpha}},
\end{equation}
with

\begin{equation}
T \equiv \frac{\alpha}{r}.
\end{equation}

We see that the main effect of aging is on the power index of the long-time tail, changing from $1 + \alpha$ for the brand new waiting time distribution density to $\alpha$ for the infinitely aged one.

For the discussion of the condition $z > 2$ we refer the readers to the excellent work of Ref. \cite{akimoto} proving that in the asymptotic time limit 
\begin{equation}
\psi(\tau, t_a) \propto \frac{t_a^{\alpha}}{\tau^{\alpha}(t_a + \tau)}.
\end{equation}
We see that aging turns $1 + \alpha$ into $\alpha$ also in this case. The aging process in this case is perennial because  $p(y,t_a)$ should get the invariant measure 
\begin{equation} \label{cheerful2}
p_{eq}(y) \propto \frac{1}{y^{z-1}}, \end{equation}
which is not normalizable. The equation of motion of Eq. (\ref{cheerful}) leaves the norm of the initial distribution $p(y, 0)$ equal to $1$ thereby leaving the time evolution towards the invariant distribution in a perennial out of equilibrium condition.  

  The survival probability corresponding to the waiting time distribution density $\psi(\tau, t_a)$ can be written under the form
\begin{equation} \label{agingsurvival}
\Psi(t, t_a) = \Psi(\tau, t_a),
\end{equation}
where $\tau = t -t_a$,
and the long-time tail with $\alpha < 1$ becomes virtually flat with no more events occurring for $t_a \rightarrow \infty$.

\section{Aging of the ML process} \label{newaging}

Let us now go back to Eq. (\ref{fundamental}) adopting a physical interpretation that establishes a connection with the ML function of Eq. (\ref{supers}). 
We study the aging survival probability defined by 

\begin{equation} \label{supertau}
F(\tau, t_a) \equiv \int_{0}^{\infty} dq p(q,t_a) exp(-q\tau),
 \end{equation}
 where $\tau = t - t_a$. 
   This is an aging survival probability of the same kind as  that of Eq. (\ref{agingsurvival}). In the case of Eq. (\ref{agingsurvival}) the choice of the coordinate $y$ at time $t_{a}$ has the effect of producing an event at time $t =t_a + \tau$, this event being the arrival of the trajectory $y(t)$ at $y = 1$. 
 In this section the event produced at $t= \tau + t_a$ is  the emission of a photon, due to the fact that $q$ is assumed to be proportional to the rate of photon emission.

 The survival probability $F(\tau,t_a)$  can be interpreted as the fluorescence intensity of a system, for instance the enzyme of Ref. \cite{fluctuatingenzyme}, with the enzyme prepared at time $t=0$ and the observation of photon emission beginning at time $t_a>0$.  For $t_a \rightarrow \infty$, $F(\tau, \infty) = E_{\alpha}(-(\lambda t)^{\alpha})$ of Eq. (\ref{supers}). In this section we show that the results of the recent work of \cite{dechant} can be used to generate through Eq. (\ref{supertau}) the long tail of the ML function. 
In Eq. (\ref{supertau}) the initial density $p(q,0)$ is 
 given by 
\begin{equation}
p(q,0) = \frac{1}{a}
\end{equation}
for $q < a$
and
\begin{equation}
p(q,0) = 0
\end{equation}
for $q \geq a$. In the large $q$ scale this initial distribution  can be interpreted 
as
\begin{equation}
p(q,0) = \delta(q). 
\end{equation}
 
  Note that the distribution density $p(q,t_a)$ is normalized. The logarithmic approximation of Eq. (\ref{short}), necessary to establish a connection with the work of Ref. \cite{dechant})  has the effect of creating a non-normalizable distribution. However, following \cite{dechant} we assume that the non-normalizable distribution density is truncated at $q = d(t_a)$ and through the time dependent normalization factor we generate the same perennial transition regime as that produced by Eq. (\ref{cheerful}) and 
 \begin{equation}
 p(0, t_a) \propto \frac{1}{t_{a}^{\delta}}
 \end{equation}
 in accordance with Eq. (\ref{important1}).

From a formal point of view Eq. (\ref{supertau}) is nothing but the Laplace
transform of $p(q,t_a)$ with respect to the variable $q$ and time $\tau$ interpreted as being 
the Laplace parameter usually called $s$.  We adopt the notation
\begin{equation}
 \hat f(s) = \mathcal{L} [f(x)].
\end{equation}
So we have
\begin{equation} \label{palatella1}
F(\tau, t_a) = \hat p(s = \tau, t_a),
\end{equation}
with
\begin{equation} \label{palatella2}
\hat p(\tau , t_a)   =  \int\limits_{0}^{\infty} \mathrm{d} q p(q,t_a) \exp(- \tau q ).
\end{equation}

Let us make the assumption that $t_a$ is finite and \emph{pdf }corresponding to the initial condition $p(q,0) = \delta(q)$ did not have enough time to become so broad as to reach the large $q$ region where 
$p(q,t)$ is expected to become closer and closer to the equilibrium distribution $\Pi(q)$ of Eq. (\ref{normalizable}). 
In this condition we have
\begin{equation}
p(q,t_a) = \frac{N(t_a)}{q^{1-\alpha}},
\end{equation}
for $q < d(t_a)$ 
and
\begin{equation}
p(q,t_a) = 0
\end{equation}
for $q \geq d(t_a)$. 
As a consequence, the normalization factor $N(t_a)$ becomes
\begin{equation}
N(t_a) = \frac{\alpha}{d^{\alpha}(t_a)}
\end{equation}
It is possible to establish the dependence of $d(t_a)$ on $t_a$, by noticing
that in accordance with Eq.(\ref{important1}) and Eq. (\ref{important2}) 
\begin{equation}\label{veryimportant}
p(a,t_{a}) = \frac{\alpha}{d^{\alpha}(t_a)} \frac{1}{a^{1- \alpha}}\propto \frac{1}{t_a^{\alpha/2}},
\end{equation}
which yields
\begin{equation}
d \propto t_{a}^{1/2}. 
\end{equation}
Using Eq. (\ref{palatella1}),  Eq. (\ref{palatella2}) and the well known 
Tauberian relation
\begin{equation}
\mathcal{L}\left\{\frac{1}{t^{\delta}}\right\}
=\left(\frac{1}{u}\right)^{1-\delta} \Gamma(1-\delta),
\end{equation}
we obtain
\begin{equation}
F(\tau, t_a) = \frac{N(t_a) \Gamma(1-\alpha)}{\tau^{\alpha}} \propto
\frac{1}{t_a^{\alpha/2}} \frac{\Gamma(1-\alpha)}{\tau^{\alpha}}, 
\end{equation}
which is the long-time limit of $E(-(\lambda \tau)^{\alpha})$ of Eq. (\ref{supers}). 

This is an interesting result. However, for very large values of $t_a$ it conflicts with the ML structure of Eq. (\ref{supers}). If we make the assumption that
\begin{equation}
F( \tau, t_a) = \left(\frac{T(t_a)}{\tau + T(t_a)}\right)^{\alpha}
\end{equation}
by assigning to $T_a$ the goal of defining the size of the short $\tau$ region,
we obtain
\begin{equation}
T(t_a) = \frac{\Gamma(1-\alpha)^{1/\alpha}}{t_{a}^{1/2}}
\end{equation}
indicating that for $t_a \rightarrow \infty$ the short time region, where the stretched exponential is expected to appear, becomes infinitesimally small.
This conflicts with the physics behind Eq. (\ref{supers}) and it is a consequence of assuming that the logarithmic idealization applies to the large- as well as to the short-$q$ region. 

The adoption of the analytical approach of the authors of Ref. \cite{dechant} give a strong support to our claim that the system evolves in time towards the equilibrium distribution density $\Pi(q)$ of Eq. (\ref{normalizable}). In fact, their analytical solution for $p(q,t)$, at $1 \gg q > a$ is proportional to $1/q^{1-\alpha}$, which fits the analytical expression of Eq. (\ref{normalizable}). At larger values of $q$ this inverse power law is truncated by a faster decay, which has the effect of creating a deviation from the ML structure of Eq. (\ref{supers}) in the short $\tau$ region, as we shall discuss in Section \ref{david}. However, in the short {q} region $p(q,t)$ reaches the power law structure necessary for properly generating the long-time tail of the ML function $E_{\alpha}(-(\lambda t)^{\alpha})$.

\section{Beyond the logarithmic approximation} \label{david}

To show the corrections to the logarithmic approximation the Langevin equation given by Eq. (\ref{fundamental1}) was integrated numerically.  The noise was chosen to be Gaussian with variance given by $\sigma^2=2D\Delta{t_a}$, where $\Delta{t_a}$ was the time step.  The deterministic part of the Langevin equation was integrated using a fourth-order Runge-Kutta algorithm.  For values of $q$ smaller than $\sqrt{2(1-\alpha)\Delta{t}}$ the deterministic part of the Langevin equation was ignored, this is equivalent to setting the potential equal to a constant in this region.  Every trajectory had its initial position at $q=1E-8$.

We compared three analytical results to the numerical ones: the waiting-time distribution for large and small waiting times, the survival probability at large values of $\tau$, and the probability distribution at large values of $t_a$.  Eq. (\ref{palatella}) yields the exponent of the power law that the waiting-time distribution is expected to exhibit at small waiting times.  

In the large $q$ limit Eq. (\ref{fundamental1}) is very well approximated by

\begin{equation}\label{largeqlangevin}\dot{q}\simeq-\frac{1+\alpha}{q}.\end{equation}  Using a procedure analogous to that adopted  to arrive at Eq. (\ref{palatella}) one obtains\begin{equation}\label{largeqmu}\mu=\frac{4+\alpha}{2}.\end{equation}  

\begin{figure}\label{Fig. 1}
\includegraphics[trim=0 390 0 125, clip=true,width=150mm]{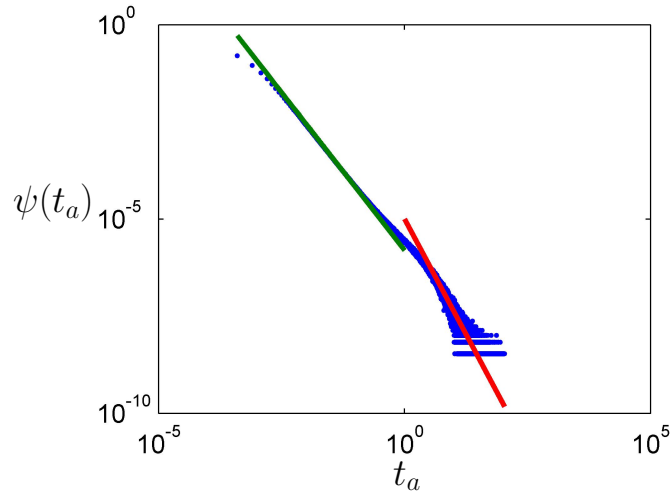}
\caption{A log-log plot of the numerical (blue) and analytical (red and green) results for the waiting-time distribution with $\alpha=0.75$.    The green(red) line indicates the power law expected for small(large) $q$.}\end{figure}  

Fig. 1 shows that the analytical prediction of Eq. (\ref{largeqmu}) is reasonably well reproduced by the numerical results, although less accurately than the analytical prediction of Eq. (\ref{palatella}). In accordance to the theory of this article, the analytical prediction of Eq. (\ref{palatella}) does not have any time limit 
in the ideal condition of a logarithmic potential. The ML potential of Eq. (\ref{potential}) is compatible with the prediction of Eq. (\ref{palatella}) in the short-time region, corresponding to the short-$q$ region, while leading in the long-time limit, corresponding to the large-$q$ region, to the power law of Eq. (\ref{largeqmu}). This result leads us to make an important remark about  $\mu = 2$. This power index that in the case of the PM map corresponds to the border between the ergodic ($\mu > 2$) and non ergodic ($\mu<2$) regime, in the case of noise-induced intermittence of this article applies to both the short- and the long-time regime. 

Fig. 2 illustrates a numerical result supporting the theoretical interpretation of Eq. (\ref{supertau}). It represents the equilibrium distribution $\Pi(q)$ of Eq. (\ref{normalizable}) obtained by integrating numerically the Langevin equation of Eq. (\ref{langevin}). This figure confirms very well that $\Pi(q)$ of Eq. (\ref{supertau}) can be interpreted as the equilibrium distribution of Eq. (\ref{supertau}) with the caution, however, of keeping in mind that this is rigorously true only for $t_a = \infty$. The condition $\alpha = 0.99$ of this figure corresponds for $E_{\alpha}(-(\lambda t)^{\alpha})$ of Eq. (\ref{supers}) to the case of a very extended stretched exponential with the inverse power law emerging at very large times. In accordance with the theoretical arguments of Section \ref{newaging} the distribution density 
for values of $q\ll1$, corresponding to ML inverse power law, and so the ML long-time regime, is realized numerically for times of the order of $t_a\approx0.1$ or less. 
At large $q$-values ($q\gtrsim1$) the distribution converges more slowly, always retaining an exponential cutoff at sufficiently large $q$, as expected from the approximate solutions in Ref. \cite{dechant} applied to the Fokker-Planck equation associated with Eq. (\ref{largeqlangevin}).

\begin{figure}\label{ensemblet2alpha99}
\includegraphics[trim=0 390 0 125, clip=true,width=150mm]{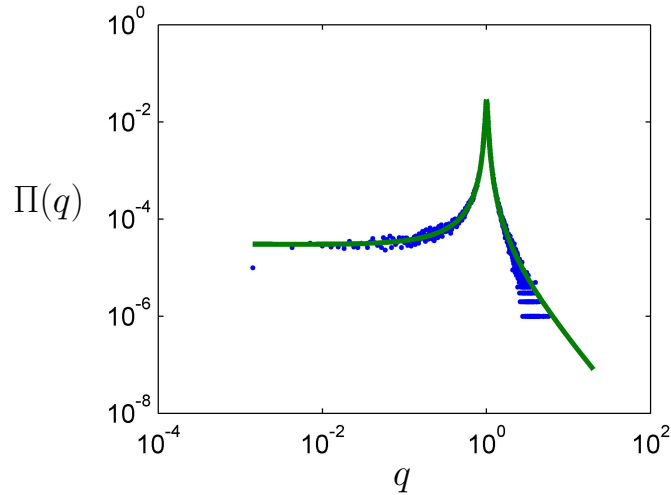}
\caption{A log-log plot of the numerical (blue) and analytical (green) probability distributions for $\alpha=0.99$.}\end{figure}

Notice that $\psi(t)$ and $\Psi(t)$ denote the waiting time distribution density between two consecutive returns to the origin of $q(t)$ driven by the Langevin Equation of Eq. (\ref{langevin}). Here we want to discuss the issue of the dynamical realization of the ML function of Eq. (\ref{supers}).  This survival probability refers to the photon emission and represents the probability that no photon is emitted up to time $t$ after the emission of the first photon at time $t = 0$. It is an idealized form of equilibrium thermodynamics in the case where the time necessary to reach equilibrium is divergent. Unfortunately we do not have the analytical theory to illustrate this survival probability at times $t_a < \infty$. We have to limit ourselves, with Fig. 3, to address this important problem using a numerical approach. Its result is shown in Fig. 3.

  \begin{figure}\label{survivalalpha99}
  \includegraphics[trim=0 390 0 125, clip=true,width=150mm]{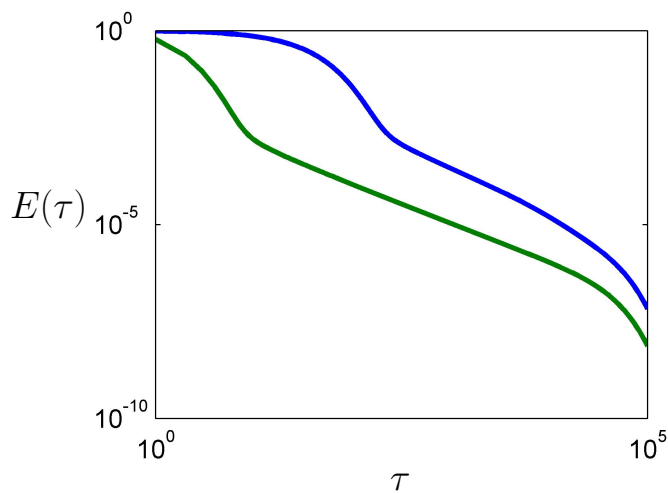}
  \caption{A log-log plot of the analytic (green) and numerical (blue) survival probabilities for $\alpha=0.99$.}\end{figure}
  
Fig. 3 shows the comparison between the numerical and theoretical survival probability for $t_a = 1$. The numerical stretched exponential is expected to tend to the theoretical stretched exponential in the limit $t_a\rightarrow\infty$. The exponential behavior at large times is an artifact of the numerical integration (equivalent to a numerical Laplace transform) applied to the dynamical distribution to obtain the survival probability.

To determine the time interval over which the small-$q$ approximation of Eq. (\ref{short}) is relevant the second moment of the distribution of trajectories generated using Eq. (\ref{fundamental1}) was compared to that obtained from using Eq. (\ref{ideal}).

\begin{figure}\label{secondmoment}
\includegraphics[trim=0 390 0 125, clip=true,width=150mm]{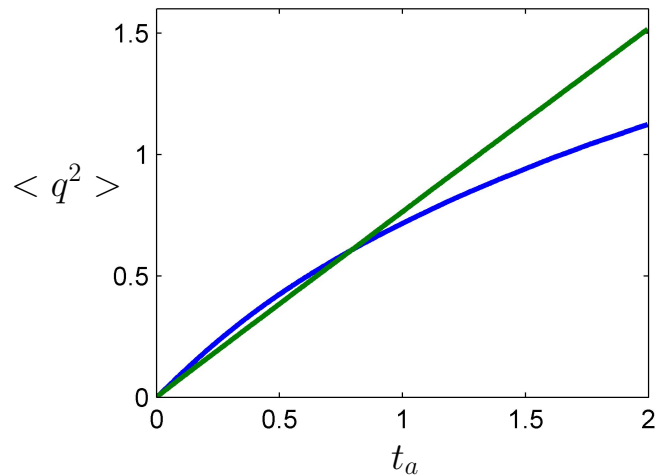}
\caption{The mean square displacement using the full Langevin equation (blue) and the logarithmic approximation (green) for $\alpha=0.75$.}\end{figure}  

It was found that the rate of change of the second moments differs significantly after a time that  decreases monotonically with increasing $\alpha$.  This is because the large-$q$ behavior of the waiting-time distribution dictates that at large times the second moment should increase relatively slowly (compared to the rate at small times), and this change in rate becomes more pronounced as $\alpha$ is increased.

\section{Concluding remarks} \label{final}

The important result of this article is the Langevin equation of Eq. (\ref{langevin}). This equation can be used as a dynamical model to explain BQD fluorescence, with $q> 0$ corresponding to the light state and $q<0$ to the dark state. 
As a benefit of this model, we get rid of the limitation to $\mu = 1.5$, in accordance with the earlier work of Ref. \cite{luigi}. In fact, we get Eq. (\ref{palatella}) making $\mu$ move from $1.5$, at $\alpha =1$ to $2$, at $\alpha = 0$. 
However, due to the corrections to the logarithmic approximation necessary to recover the physical condition of Ref. \cite{luigi}, the resulting process is multi scaling and it is rigorously mono scaling only for $\alpha = 0$, yielding $\mu = 2$. 

The Langevin equation of Eq. (\ref{langevin}) can also be used to predict an anomalous form of fluorescence, with a survival probability generating an aging process completely different from that usually associated to the distribution density of $\psi^{BQD}(\tau)$. If we limit ourselves to the short-time limit, this form of aging is explained using  the weak-chaos induced intermittence of Section \ref{pm}, and consequently by means of Eq. (\ref{cheerful}). Actually, the aging process of the fluorescence process of Eq. (\ref{supertau}) has the effect of affecting the short- rather than the long-time limit of the corresponding survival probability.  The emergence of $\mu \leq 2$ corresponds to an out of equilibrium process that in the case of weak chaos corresponds to $p(y,t_a)$ of Eq. (\ref{cheerful}) in a perennial motion toward equilibrium. The new form of aging corresponds to $p(y,t_a)$ of Eq. (\ref{supertau}) moving towards the equilibrium distribution of Eq. (\ref{normalizable}). In apparent conflict with the theoretical interpretation of aging,  advocated by the authors of Ref. \cite{korabel1,korabel2,akimoto}),  in terms of non-normalizable density measures, this equilibrium is normalizable, but all its moments $q^{m}$ with $m>0$ are divergent. In practice, this corresponds to a perennial out of equilibrium condition.

\emph{Acknowledgements}
D.L and P.G.  warmly thank Welch and ARO for their support through Grants No. B-1577 and No. W911NF-11-1-0478, respectively.

.

\end{document}